\documentclass[aps,pre,amssymb,amsfonts,a4paper,twoside,twocolumn]{revtex4-1}
\usepackage{amssymb,amsmath,bm}
\usepackage{graphicx}
\usepackage{enumerate}
\usepackage[utf8]{inputenc}
\usepackage{cancel}
\usepackage{color}
\usepackage[colorlinks,linkcolor=blue,urlcolor=blue,citecolor=blue]{hyperref}
\usepackage{float}
\usepackage{soul}
\begin{document}

\title{Tighter thermodynamic bound on speed limit in systems with unidirectional transitions}
\author{Deepak Gupta$^1$} \email{Corresponding author: phydeepak.gupta@gmail.com} \author{Daniel M. Busiello$^2$}\email{daniel.busiello@epfl.ch}
\affiliation{$^1$Dipartimento di Fisica `G. Galilei', INFN, Universit\`a di Padova, Via Marzolo 8, 35131 Padova, Italy}
\affiliation{$^2$Ecole Polytechnique F\'ed\'erale de Lausanne (EPFL), Institute of Physics Laboratory of Statistical Biophysics, 1015 Lausanne, Switzerland}

\date{\today}
\begin{abstract}
We consider a general discrete state-space system with both unidirectional and bidirectional links. In contrast to bidirectional links, there is no reverse transition along the unidirectional links. Herein, we first compute the statistical length and the thermodynamic cost function for transitions in the probability space, highlighting contributions from total, environmental, and resetting (unidirectional) entropy production. Then, we derive the thermodynamic bound on the speed limit to connect two distributions separated by a finite time, showing the effect of the presence of unidirectional transitions. Novel uncertainty relationships can be found for the \textit{temporal} first and second moments of the average resetting entropy production. We derive simple expressions in the limit of slow unidirectional transition rates. Finally, we present a refinement of the thermodynamic bound, by means of an optimization procedure. We numerically investigate these results on systems that stochastically reset with constant and periodic resetting rate.
\end{abstract}
\maketitle

\section{Introduction}
\label{info}
Small systems coupled with heat baths evolve under stochastic dynamics. Their probabilistic description at all time can be understood using the master equation or the Fokker-Planck equation \cite{vK}. In this kind of systems, an increasing interest has been devoted in studying thermodynamic observables, such as heat dissipated in the environment, work done on the system, entropy production, probability currents, and so on. The estimation of these fluctuating quantities requires a description at the level of a single stochastic trajectory, which can be obtained within the celebrated framework of {\it stochastic thermodynamics} \cite{ST-0,ST}. Interestingly, these stochastic quantities follow some universal results in the non-equilibrium physics; namely, fluctuation theorem \cite{US}, Jarzynski equality \cite{JE}, Crooks fluctuation relation \cite{CR}, and thermodynamic uncertainty relations (TURs) \cite{tur-0}.

More recently, among those various stochastic systems, those exhibiting {\it unidirectional transitions} have attracted attention due to broad interest in several areas of science. Examples include spontaneous decay of atom \cite{rahav}, directed percolation in liquid crystal \cite{Masaki}, TASEP \cite{Ronald,saha}, driven inelastic Lorentz-gas \cite{Chong}, and stochastic resetting \cite{Evans_2011} (see also \cite{Evans_2020} and references therein). In these systems, there is no reverse transition along the unidirectional link, in stark contrast with the case of bidirectional transitions. In this direction, only few attempts have been made in understanding the thermodynamics of these systems; namely, the first and second law of thermodynamics  \cite{Fuchs_2016, Busiello_2020}, work fluctuation and Jarzynski equality \cite{Gupta_2020}, integral fluctuation theorem \cite{Pal_2017}.

Information theory provides a fascinating language to rephrase most of the well-known results in thermodynamics, and to shed new light on their meaning. The connection between information and thermodynamics goes back to the seminal work by Maxwell and his thought experiment \cite{max}, and has been recently properly presented by the work by Parrondo {\it et al.} \cite{Parrondo}. The intense research activity on this topic has led to numerous remarkable findings, e.g., thermodynamic uncertainty relations \cite{tur-1,tur-2,tur-3,tur-4}, a connection of entropy and K-L divergence \cite{footprints}, the convex property of relative entropy \cite{esposito}, the generalized Jarzynski equality \cite{sagawa}, a bound on the entropy production due to information flow  \cite{causal}. Several interesting results in close to equilibrium systems have been also obtained following differential geometric interpretation of thermodynamics \cite{neq-1,neq-2,neq-3,neq-4,crooks,sivak,sivak-2}. Arbitrarily far from equilibrium, Ito \cite{ito} found a thermodynamic bound on the time to reach a final state from an initial state, strengthening the link between information and thermodynamics. Recently, this result have been used to study the relation between the cost in the bacterial growth and the evolution time \cite{bacteria}, and between adaptation speed and its thermodynamic cost for {\it E. coli} \cite{ecoli}. Similarly, with different approaches, speed limit has also been investigated in Refs. \cite{Funo,vo}.  Here, we aim to extend the formalism and the results presented in Ref.~\cite{ito} to systems with unidirectional transitions. 

In this paper, we consider a discrete state-space system with both unidirectional and bidirectional transitions. In this system, we first compute the different contributions of entropy productions. We remark that the analysis is based on the recent work by Ito \cite{ito}, and it is carried out in the case of a system with stochastic resetting and for the general case of multiple unidirectional transitions. In what follows, we will use the name \textit{resetting entropy production to identify the contribution stemming from unidirectional links only.} Then, we compute the {\it thermodynamic action} or the {\it thermodynamic cost function} and the {\it statistical length} \cite{crooks} for the system to make transition from an initial state to the final one in a finite time. We show that the time to connect these two distributions is bounded by a function which contains the information of thermodynamic cost function and the statistical length. Here, we obtain three important results: 1) the thermodynamic cost function and the thermodynamic bound on the speed limit to connect the initial and final distributions for systems with both unidirectional and bidirectional transitions, 2) two inequalities and an uncertainty relation (TUR-like) involving resetting entropy production, and 3) a novel bound on the speed limit which is stronger than that obtained in \cite{ito}. Further, we stress that such a tighter bound is also applicable to systems with only bidirectional transitions.

The rest of the paper is organized as follows. In Sec.~\ref{setup}, we describe a system with stochastic resetting and compute the entropy productions associated with it. In order to understand the distance between two distributions, we discuss K-L divergence in Sec. \ref{sec-kl-div}. Sec.~\ref{stlen-ctfun} contains the definition of the statistical length and the thermodynamic action, and their connection with thermodynamic quantities. Using these ingredients, we discuss the thermodynamic bound on the speed limit to connect two distributions in Sec.~\ref{unid}, in the presence of unidirectional transition rates. Moreover, we investigate the effect of small resetting rate on the bound. Sec.~\ref{TUR} is dedicated to novel inequalities and an uncertainty relation (TUR-like) for the resetting entropy production. In Sec.~\ref{speed}, we propose a derivation of a {\it new} bound on the speed limit, and that can also be made tighter using an optimization scheme (in Sec.~\ref{opt}). Finally, we summarize our paper in Sec.~\ref{summ}. In Appendix \ref{ul-sec} and \ref{FIM}, we discuss, respectively, thermodynamics and Fisher information metric of systems with multiple unidirectional links.

\section{Set up}
\label{setup}
\begin{figure}
  \begin{center}
  \includegraphics[width=0.8 \columnwidth]{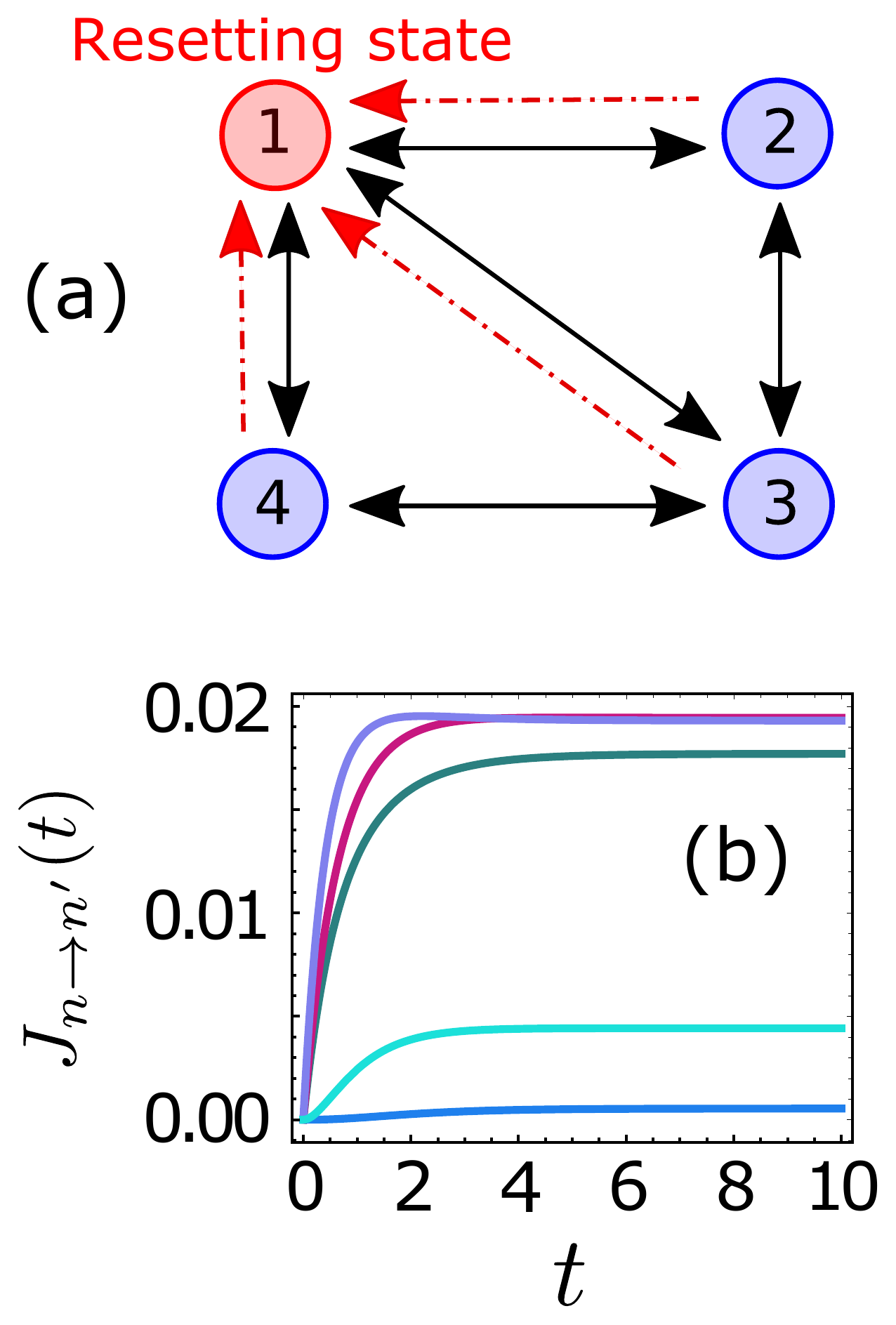}
    \caption{Thermodynamic flux \eqref{cj} across $(n,n^\prime)$ link with respect to observation time $t$. Herein, we consider a network of four states [panel (a)] with an initial condition $(p_1(0), p_2(0), p_3(0), p_4(0))=(0.4, 0.2, 0.25, 0.15)$. We have taken bidirectional transition rates (solid arrows) $W_{2\to 1}=0.5, W_{3 \to 2}=0.1, W_{3 \to 1}=0.5, W_{4 \to 3}=0.5, W_{1 \to 4}=0.5$, and the reverse transition rates are fixed by the detailed balanced condition at time $t=0$. Moreover, we consider resetting rates $\gamma_n=0.1$ for all $n$ (dot-dashed arrows). Clearly, each thermodynamic flux reaches to a non-zero stationary value at large time indicating the system is in a non-equilibrium steady state [panel (b)].}
    \label{fig:flux}
  \end{center}
\end{figure}

Consider a system composed of $N$ discrete states. Suppose the system jumps from a state $n^\prime$ to another state $n$ with a non-zero transition rate $W_{n^\prime\to n}>0$, and there also exists a non-zero reverse transition $W_{n\to n^\prime}>0$. In addition to this, the system stochastically resets to a given state with a rate $\gamma_n>0$ from a state $n$. Notice that unlike $W_{n\to m}$, corresponding to each $\gamma_n$ there do not exist a reverse transition. Thus, the evolution of the probability of the system to be in state $n$ at time $t$ is described by the following master equation \cite{vK}:     
\begin{align}
&\dfrac{\mathrm{d}p_n(t)}{\mathrm{d}t}=\sum_{n^\prime\neq n}^{N}[W_{n^\prime \to n}~p_{n^\prime}(t)-W_{n\to n^\prime }~p_n(t)]~+\nonumber\\
&-\gamma_n~p_n(t)+\delta_{n,1}\sum_{m=1}^{N}\gamma_m~p_m(t),\qquad 1\leq n\leq N,
\label{dyna-1}
\end{align}
where the first term on the right-hand side corresponds to the bidirectional transitions. The second and third terms on the right-hand side of Eq. \eqref{dyna-1}, respectively, account for the contributions due to the loss in the probability of state $n$ and the gain in the probability of state $1$ (i.e., the resetting state) from all states. In the above equation \eqref{dyna-1}, $\delta_{n,m}$ is the Kronecker delta defined as $\delta_{n,m}=1$ for $n=m$, and $0$ otherwise. Since the probability distribution over all states has to be normalized, $\sum_{n=1}^{N}~p_n(t)=1$, the sum of the bidirectional transition rates from a given states to all states is zero: $\sum_{n^\prime=1}^{N}~W_{n\to n^\prime}=0$. This implies that $W_{n\to n}=-\sum_{n^\prime\neq n}W_{n\to n^\prime}<0$. 
A schematic diagram of a four states system is shown in Fig.~\ref{fig:flux}(upper panel), where solid arrows are the bidirectional links and the dot dashed ones are the unidirectional (resetting) links. 

In general, the transition rates on the right-hand side of Eq.~\eqref{dyna-1} may depend on time $t$. In what follows, unless specified, the transition rates have an explicit time-dependence. Note that in Eq.~\eqref{dyna-1}, we are considering systems that resets to only one state. Nevertheless, the generalization for a system which resets stochastically to any states is straightforward to handle (see Appendix \ref{ul-sec}), i.e., there can be multiple unidirectional links in the network \cite{Busiello_2020}.

The average entropy of the system is given by 
\begin{align}
S^{\mathrm{sys}}=-\sum_{n=1}^{N}p_n(t) \ln p_n(t),
\label{sys-ent}
\end{align}
where $p_n(t)$ is the solution of master equation \eqref{dyna-1} subject to an initial condition $p_n(0)$. In the above equation \eqref{sys-ent}, we have set the Boltzmann's constant $k_B$ equal to 1. For convenience, in the following, we will not explicitly write  the  time-dependence in $p_n(t)$.

Differentiating Eq.~\eqref{sys-ent} with respect to time $t$, we get the average entropy production of the system
\begin{align}
\dot S^{\mathrm{sys}}=-\sum_{n=1}^{N}\dot p_n \ln p_n-\sum_{n=1}^{N} \dot p_n,
\label{dotsys}
\end{align} 
where the dot indicates a time derivative. In the expression above, Eq.~\eqref{dotsys}, the last term is equal to zero since the probability distribution is normalized.

Substituting Eq. \eqref{dyna-1} in Eq.~\eqref{dotsys}, we get
\begin{align}
\dot S^{\mathrm{sys}}&=-\sum_{n=1}^{N} \ln p_n \bigg[ \sum_{n^\prime\neq n}(W_{n^\prime \to n}~p_{n^\prime}-W_{n\to n^\prime }~p_n)~+\nonumber\\&~~~~~~~~~~~~~~~~~~~~~~~~-\gamma_n~p_n
+\delta_{n,1}\sum_{m=1}^{N}\gamma_m~p_m\bigg]\nonumber\\
&=\sum_{n^\prime, n} W_{n^\prime \to n}~p_{n^\prime} \ln \dfrac{p_{n^\prime}}{p_{n}}+\sum_n \gamma_n~p_n\ln \dfrac{p_{n}}{p_{1}},
\label{eq1}
\end{align}
where $1\leq n,n^\prime\leq N$. Here, the second equality is obtained from the first one using that $W_{n\to n}=-\sum_{n^\prime\neq n}W_{n\to n^\prime}$. Following \cite{Busiello_2020}, the terms on the right-hand side of Eq.~\eqref{eq1} can be rewritten as following:
\begin{align}
\dot S^{\mathrm{sys}}=\overbrace{\dfrac{1}{2}\sum_{n^\prime, n} J_{n^\prime\to n}~F_{n^\prime\to n}}^{\dot S^{\mathrm{tot}}}-\overbrace{\dfrac{1}{2}\sum_{n^\prime, n} J_{n^\prime\to n} \sigma^{\mathrm{bath}}_{n^\prime\to n}}^{\dot S^{\mathrm{bath}}}~+\nonumber\\
 -\overbrace{\sum_n J^{\mathrm{reset}}_n \sigma_n^{\mathrm{reset}}}^{\dot S^{\mathrm{reset}}},
\label{eq2}
\end{align}
where $\dot S^{\mathrm{tot}}$, $\dot S^{\mathrm{bath}}$, and $\dot S^{\mathrm{reset}}$, respectively, are the total entropy production, the bath entropy production, and the resetting entropy production \cite{Fuchs_2016,Busiello_2020}. Herein, the factor $1/2$ avoids the double-counting in the first two summations. Notice that when the system has multiple unidirectional links along with bidirectional ones, the resetting entropy production (which will be recognized as the unidirectional entropy production) can be easily modified (see Appendix \ref{ul-sec}) \cite{Busiello_2020}. In the above equation \eqref{eq2}, we identify the following terms: 
\begin{align}
J_{n^\prime\to n}&=W_{n^\prime \to n}~p_{n^\prime}-W_{n\to n^\prime }~p_n,\label{cj}\\
F_{n^\prime\to n}&=\ln \dfrac{W_{n^\prime \to n}~p_{n^\prime}}{W_{n\to n^\prime }~p_n}\label{fj},\\
\sigma^{\mathrm{bath}}_{n^\prime\to n}&=\ln \dfrac{W_{n^\prime \to n}}{W_{n\to n^\prime }}\label{sgj},\\
\sigma_n^{\mathrm{reset}}&=\ln \dfrac{p_{1}}{p_{n}}\label{s-reset}\\
J^{\mathrm{reset}}_{n}&=\gamma_n~p_n\label{rj}.
\end{align}
These represent, respectively, the thermodynamic flux \eqref{cj} for the transition $n^\prime\to n$ and the corresponding thermodynamic force \eqref{fj}, the change in entropy in the bath due to transition from $n^\prime\to n$, the change in resetting entropy \eqref{s-reset} and the flux due to resetting the system from state $n$ to 1 \eqref{rj}. Notice that both the thermodynamic force and the change in entropy in the bath are anti-symmetric under exchange of indices, i.e., $F_{n^\prime\to n}=-F_{n\to n^\prime}$ and $\sigma^{\mathrm{bath}}_{n^\prime\to n}=-\sigma^{\mathrm{bath}}_{n\to n^\prime}$.

When the system has time-independent rates and there is no resetting mechanism (i.e., $\gamma_n=0$ for all $n$), it eventually relaxes to a stationary state at large time. This state can be either equilibrium or non-equilibrium stationary state depending on whether or not, respectively, the transition rates obey the detailed balance: $W_{n^\prime \to n}~p_{n^\prime}=W_{n\to n^\prime }~p_n$ for each $(n,n^\prime)$ link. When the detailed balance condition is not satisfied, the system is characterized by a total entropy production, $\dot S^{\mathrm{tot}}$. Furthermore, in the presence of time-independent resetting rate $\gamma_n$ (see Fig.~\ref{fig:flux}), the system reaches a non-equilibrium steady state even when bidirectional transitions obey detailed balance at time $t=0$: $W_{n^\prime \to n}~p_{n^\prime}(0)=W_{n\to n^\prime }~p_n(0)$ for each $(n,n^\prime)$ link.

\section{K-L divergence and the Fisher metric}
\label{sec-kl-div}
Consider two distributions $\vec P=(p_1,p_2,\dots,p_N)$ and $\vec Q=(q_1,q_2,\dots,q_N)$, where $p_n\geq 0$ and $q_n\geq 0$, for $n = 1,\dots, N$. These distributions are normalized: $\sum_n p_n=1$ and $\sum_n q_n=1$. The distance between $\vec P$ and $\vec Q$ is given by the Kullback-Leibler (K-L) divergence \cite{Cover}:
\begin{align}
D_{\mathrm{KL}}(\vec P||\vec Q)=\sum_{n=1}^{N} p_n \ln \dfrac{p_n}{q_n}.
\end{align} 
Let us first summarize the properties of this measure: 1) it is not a symmetric distance, i.e., $D_{\mathrm{KL}}(\vec P||\vec Q)\neq D_{\mathrm{KL}}(\vec Q||\vec P)$, 2) it does not follow triangle inequality, and 3) it is non-negative: $D_{\mathrm{KL}}(\vec P||\vec Q)\geq0$, where equality holds if and only if $\vec P=\vec Q$. 

Now suppose that the distributions $\vec P$ and $\vec Q$ differ by a small amount, i.e., $\vec Q=\vec P+\vec{\mathrm{d}P}$. Therefore, the distance between these two distributions would be  
\begin{align}
D_{\mathrm{KL}}(\vec P||\vec P+\vec{\mathrm{d}P})=\sum_{n=1}^{N} p_n \ln \dfrac{p_n}{p_n+\mathrm{d}p_n}.
\end{align} 
Expanding the right-hand side of the above equation up to second order, we get
\begin{align}
D_{\mathrm{KL}}(\vec P||\vec P+\vec{\mathrm{d}P})\approx\dfrac{1}{2}\sum_{n=1}^{N} \dfrac{(\mathrm{d}p_n)^2}{p_n},
\label{kl-div}
\end{align}
where we have used the condition $\sum_{n=1}^{N} \mathrm{d}p_n=0$. From the right-hand side of the above equation, we can identify the square of a line element \cite{ito}: 
\begin{align}
\mathrm{d}s^2=\sum_{n=1}^{N} \dfrac{(\mathrm{d}p_n)^2}{p_n}\approx 2~D_{\mathrm{KL}}(\vec P||\vec P + \vec{\mathrm{d}P}),
\end{align}
If the system can be externally modulated using a set of protocols $\vec \lambda=(\lambda_1,\lambda_2,\dots, \lambda_m)$, then we write the above equation as
\begin{align}
\mathrm{d}s^2=\sum_{n=1}^{N}~p_n \bigg(\sum_{i=1}^{m}\dfrac{\mathrm{\partial}\ln p_n}{\mathrm{\partial}\lambda_i}~\mathrm{d}\lambda_i\bigg)^2\nonumber\\
=\sum_{i,j} \mathrm{d}\lambda_i~\mathrm{d}\lambda_j~g_{i,j},
\label{fmat}
\end{align}
where $g_{i,j}=\sum_{n=1}^N p_n \frac{\partial \ln p_n} {\partial \lambda_i}\frac{\partial \ln p_n} {\partial \lambda_j}$ is the Fisher's information metric \cite{Cover}.  In the following, we use the quantity ${\rm d}s$ to compute statistical length and thermodynamic cost function \cite{ito}.

\section{Statistical length and thermodynamic cost function}
\label{stlen-ctfun}
Let us consider a system whose evolution is characterized by only one parameter, that is the time over which we observe its evolution, i.e., $\lambda=t$. Our interest is to compute how far is the final distribution at $t=\tau$ from the initial distribution at $t=0$, given that the evolution of the system is governed by the Eq.~\eqref{dyna-1} subjected to initial condition $\vec P(t=0)=[p_1(0), p_2(0), \dots, p_N(0)]$. A measure of this distance that takes into account the whole dynamics, not just initial and final states, is provided by the statistical length $\mathcal{L}$, defined as \cite{crooks,ito} 
\begin{align}
\mathcal{L}=\int ~\mathrm{d}s=\int_0^\tau~\mathrm{d}t~\dfrac{\mathrm{d}s}{\mathrm{d}t}.
\label{stln}
\end{align}
Here, $\frac{\mathrm{d}s}{\mathrm{d}t}$ is the intrinsic speed which the probability vector traverses the path with \cite{PRX}. Stated differently, the statistical length $\mathcal{L}$ measures the total distance of the path covered by the probability vector from time $t=0$ to $t=\tau$, i.e., from an initial distribution to a final distribution, in the probability space.

Similarly, we can define a measure to quantify the cost of such a dynamical evolution. This is the thermodynamic action or thermodynamic cost function defined as \cite{crooks,ito}
 \begin{align}
\mathcal{C}=\dfrac{1}{2}\int_0^\tau~\mathrm{d}t~\dfrac{\mathrm{d}s^2}{\mathrm{d}t^2}.
\label{costfn}
\end{align}

In order to derive an explicit expression for the thermodynamic cost to connect two distributions, we first compute the integrand in Eq.~\eqref{costfn}. Using the definition of Fisher's metric for one parameter $\lambda = t$ [see Eq. \eqref{fmat}], we write 
\begin{align}
\dfrac{\mathrm{d}s^2}{\mathrm{d}t^2}&=\sum_{n=1}^{N} p_n \bigg(\dfrac{\mathrm{d}\ln p_n}{\mathrm{d}t}\bigg)^2\nonumber\\
&=-\sum_{n=1}^{N} p_n \dfrac{\mathrm{d}}{\mathrm{d}t}\bigg(\dfrac{1}{p_n}\dfrac{\mathrm{d} p_n}{\mathrm{d}t}\bigg).
\label{ds2-def}
\end{align}

Now, we substitute the value of terms inside the parenthesis using Eqs. \eqref{dyna-1} and \eqref{fj}. Therefore, we get
\begin{align}
\dfrac{\mathrm{d}s^2}{\mathrm{d}t^2}&=-\sum_{n=1}^{N} p_n \dfrac{\mathrm{d}}{\mathrm{d}t}\bigg[\sum_{n^\prime\neq n} W_{n\to n^\prime}(e^{-F_{n^\prime\to n}}-1)-\gamma_n+\nonumber\\
&~~~~~~~~~~~~~~~~~~~~~~~~~~~~~~~~+\dfrac{\delta_{n,1}}{p_n}\sum_{m=1}^{N}\gamma_m p_m\bigg],
\end{align}
Further, using $F_{n\to n}=0$ and $W_{n\to n}=-\sum_{n^\prime\neq n}W_{n\to n^\prime}$ in the above equation, we rewrite
\begin{align}
\dfrac{\mathrm{d}s^2}{\mathrm{d}t^2}&=-\sum_{n=1}^{N} p_n \dfrac{\mathrm{d}}{\mathrm{d}t}\bigg[\sum_{n^\prime} W_{n\to n^\prime}e^{-F_{n^\prime\to n}}-\gamma_n+\nonumber\\
&~~~~~~~~~~~~~~~~~~~~~~~~~~~~~~~~+\dfrac{\delta_{n,1}}{p_n}\sum_{m=1}^{N}\gamma_m p_m\bigg].
\label{ds2dt2}
\end{align}
\begin{figure*}
    \centering
    \includegraphics[width=2 \columnwidth]{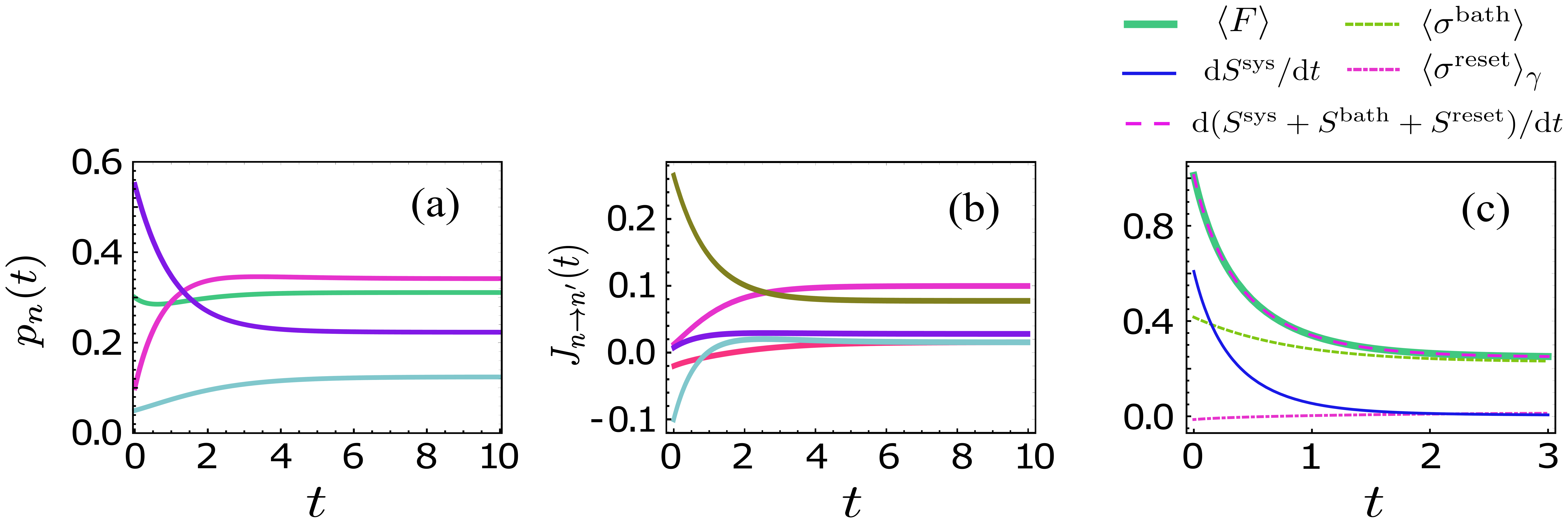}
    \caption{Probability distribution, thermodynamic flux, and entropy production with respect to time $t$. We consider a network of four states as shown in Fig.~\ref{fig:flux}(upper panel) with transitions rates $W_{2 \to 1}=0.5, W_{1 \to 2}=0.3~W_{2 \to 1}, W_{3 \to 2}=0.1, W_{2 \to 3}=0.5~W_{3 \to 2}, W_{3 \to 1}=W_{1 \to 3}=0.5, W_{4 \to 3}=0.5, W_{3 \to 4}=0.2~W_{4 \to 3}, W_{1 \to 4}=0.5, W_{4 \to 1}=0.5~W_{1 \to 4}$, $\gamma_n=0.1$ for all $n$, and the system is initialize as $(p_1(0), p_2(0), p_3(0), p_4(0))=(0.3, 0.05, 0.1, 0.55)$. In panel (a), we show the evolution of the system, and the thermodynamic flux \eqref{fj} across each link is displayed with the observation time  in panel (b). In panel (c), we show the average entropy productions with respect to time, where the angular brackets indicate the averaging as explained in the text. Here, we can see that $\langle F \rangle$ (thick green curve) is the average total entropy production [i.e., the first term on the right-hand side of Eq.~\eqref{eq2}] which is the sum of three contributions: 1) the average entropy production of system, 2) the average entropy production in bath, and 3) the average entropy production due to resetting [see Eq.~\eqref{eq2}].}
    \label{fig:reset-1}
\end{figure*}

In the following, we compute each term on the right-hand side of Eq.~\eqref{ds2dt2}. Let us first consider the first term on the right-hand side of Eq. \eqref{ds2dt2}:
\begin{align}
\sum_{n,n^\prime} p_n \dfrac{\mathrm{d}}{\mathrm{d}t}\bigg[-W_{n\to n^\prime}e^{-F_{n\to n^\prime}}\bigg]&= -\sum_{n,n^\prime} p_n \bigg[\dfrac{\mathrm{d}W_{n\to n^\prime}}{\mathrm{d}t}e^{-F_{n\to n^\prime}}\nonumber\\
&-W_{n\to n^\prime}e^{-F_{n\to n^\prime}} \dfrac{\mathrm{d}F_{n\to n^\prime}}{\mathrm{d}t}\bigg].
\end{align}
In order to proceed forward, we substitute $F_{n\to n^\prime}$ from Eq. \eqref{fj}, and then, we employ the anti-symmetric property of both thermodynamic force and entropy change in the bath due to transitions along bidirectional links. Finally, we get
\begin{align}
\sum_{n,n^\prime} p_n \dfrac{\mathrm{d}}{\mathrm{d}t}\bigg[-W_{n\to n^\prime}e^{-F_{n\to n^\prime}}\bigg]=\bigg\langle\dfrac{\mathrm{d}\sigma^{\mathrm{bath}}}{\mathrm{d}t}\bigg\rangle-\bigg\langle\dfrac{\mathrm{d}F}{\mathrm{d}t}\bigg\rangle.
\label{fpart}
\end{align}
Here the angular brackets indicate the average over all possible trajectories, defined as $\langle \Omega\rangle=(1/2)\sum_{n,m}\Omega_{n\to m}J_{n\to m}$ \cite{Busiello_2020}. Similarly, by computing the last two terms on the right-hand side of Eq.~\eqref{ds2dt2}, we obtain
\begin{align}
\sum_{n=1}^{N} p_n \bigg[\dfrac{\mathrm{d}\gamma_n}{\mathrm{d}t}-\dfrac{\mathrm{d}}{\mathrm{d}t}&\bigg\{\dfrac{\delta_{n,1}}{p_n}\sum_{m=1}^{N}\gamma_m p_m\bigg\}\bigg]=\nonumber\\
&=\sum_{n=1}^{N}\bigg[ \dfrac{\gamma_n p_n}{p_1}\dfrac{\mathrm{d}p_1}{\mathrm{d}t}-\gamma_n \dfrac{\mathrm{d}p_n}{\mathrm{d}t} \bigg]\nonumber\\
&=\sum_{n=1}^{N} \gamma_n p_n \dfrac{\mathrm{d}}{\mathrm{d}t} \bigg(\ln \dfrac{p_1}{p_n}\bigg)\nonumber\\
&=\bigg\langle\dfrac{\mathrm{d}\sigma^{\mathrm{reset}}}{\mathrm{d}t}\bigg\rangle_\gamma.
\label{r-term}
\end{align}
Here the angular brackets with the subscript $\gamma$ indicate the average over all possible resetting steps in a trajectory, which is defined as $\langle \omega\rangle_\gamma=\sum_{n}\omega_{n}J_n^{\mathrm{reset}}$. The term $1/2$ is missing since in this case each transition rate is counted once by construction.

Therefore, using Eqs. \eqref{fpart} and \eqref{r-term}, we finally obtain 
\begin{align}
\dfrac{\mathrm{d}s^2}{\mathrm{d}t^2}&=\bigg\langle\dfrac{\mathrm{d}\sigma^{\mathrm{bath}}}{\mathrm{d}t}\bigg\rangle-\bigg\langle\dfrac{\mathrm{d}F}{\mathrm{d}t}\bigg\rangle+\bigg\langle\dfrac{\mathrm{d}\sigma^{\mathrm{reset}}}{\mathrm{d}t}\bigg\rangle_\gamma.
\label{fexp-ds2}
\end{align}
Similarly, the above result can also be generalized for a system with multiple unidirectional system (see Appendix \ref{FIM}).
Substituting the above equation in Eqs.~\eqref{stln} and \eqref{costfn}, we obtain, respectively, the expression for the statistical length $\mathcal{L}$ and the thermodynamic cost function $\mathcal{C}$, in terms of the entropic contributions in the stochastic resetting system.

In Fig.~\ref{fig:reset-1}(a), we numerically evolve the dynamical equations \eqref{dyna-1} for the network shown in Fig.~\ref{fig:flux}(upper panel) and show the probability to be in each state with respect to time. Herein, we choose all transitions rates to be time-independent. Consequently, the system reaches a non-equilibrium steady state [see Fig.~\ref{fig:reset-1}(b)] as time progresses and the system entropy production \eqref{dotsys} vanishes in the non-equilibrium steady state [see thin solid blue curve in Fig.~\ref{fig:reset-1}(c)]. Moreover, we can see that $\langle F \rangle \to \langle \sigma^{\mathrm{bath}}\rangle +\langle \sigma^{\mathrm{reset}} \rangle_\gamma$ as expected, in the large time limit.

In the next section, we build a relation between the thermodynamic cost function \eqref{costfn} and the statistical length \eqref{stln}.

\section{The role of unidirectional transitions}
\label{unid}

Here, we first show that the relation between $\mathcal{L}$ and $\mathcal{C}$ can be shaped in the form of a thermodynamic bound for the speed at which the system can go from an initial to a final distribution. In order to do so, we employ the Cauchy-Schwartz inequality for two real functions $f(t)$ and $g(t)$:
\begin{align}
&\bigg[\int_0^\tau\mathrm{d}t~f(t)~g(t)\bigg]^2 \leq \int_0^\tau\mathrm{d}t_1~f^2(t_1)\int_0^\tau\mathrm{d}t_2~g^2(t_2).
\label{cauchy}
\end{align}

Now we substitute $g(t)=\phi$ (an arbitrary constant $\in \mathbb{R}$) and $f(t)=\frac{\mathrm{d}s}{\mathrm{d}t}$ in Eq.~\eqref{cauchy}, and we obtain [see Eqs. \eqref{stln} and \eqref{costfn}]:
\begin{align}
\mathcal{C}_\tau\geq\dfrac{\mathcal{L}^2}{2\tau^2},
\label{cst-1}
\end{align}
where $\mathcal{C}_\tau=\frac{\mathcal{C}}{\tau}$ is a scaled quantity, and the term on the right-hand side can be interpreted as the square of the speed at which the probability vector traces a path to connect initial and the final states.

Rearranging the above inequality \eqref{cst-1}, the bound on the speed limit $\tau$ can be expressed as:
\begin{align}
\tau \geq \dfrac{\mathcal{L}}{\sqrt{2 \mathcal{C}_\tau}}.
\label{tur-cauchy}
\end{align}
We stress that although the form of the thermodynamic bound remains same as in Ref.~\cite{ito} even for the case of system with unidirectional transition, the difference lies in entropic contributions in both $\mathcal{L}$ and $\mathcal{C}$.

In order to show how the presence of unidirectional transitions modifies the bound on the speed limit, let us start considering a network without unidirectional links. It will satisfy the inequality in Eq.~\eqref{tur-cauchy}, with:
\begin{equation}
\frac{{\rm d}s^2}{{\rm d}t^2} = \left\langle \frac{{\rm d}\sigma^{\rm bath}}{{\rm d}t} \right\rangle - \left\langle \frac{{\rm d}F}{{\rm d}t} \right\rangle = \frac{{\rm d}s^2}{{\rm d}t^2}\bigg|_0.
\end{equation}
We name the bound on the speed limit for this network, i.e,. the right-hand side of Eq.~\eqref{tur-cauchy}, $\tau_0^{\rm bound}$. 

Consider, now, the same network with an additional time-independent unidirectional transition rate, $\gamma$, from the node $n_1$ to the node $n_2$. The bound maintains the same form, but the square of the intrinsic speed becomes equal to the one reported in Eq.~\eqref{fexp-ds2}. In order to study what happens in the limit of $\gamma$ much smaller than all other transition rates, i.e., slow resetting, we expand the statistical length up to the first order:
\begin{gather}
\mathcal{L} \approx \int_0^\tau dt \left( \sqrt{\frac{{\rm d}s^2}{{\rm d}t^2}\bigg|_0} + \frac{\left\langle \frac{{\rm d}\sigma^{\rm reset}}{{\rm d}t} \right\rangle_\gamma}{2 \sqrt{\frac{{\rm d}s^2}{{\rm d}t^2}\big|_0}} \right) \approx \mathcal{L}^{(0)} + \mathcal{L}^{(\gamma)}.\label{l-exp}
\end{gather}
The cost function can be rewritten as follows:
\begin{align}\mathcal{C}_\tau = \frac{1}{2 \tau} \int_0^\tau dt \left( \frac{{\rm d}s^2}{{\rm d}t^2}\bigg|_0 + \left\langle \frac{{\rm d}\sigma^{\rm reset}}{{\rm d}t} \right\rangle_\gamma \right) = \mathcal{C}_\tau^{(0)} + \mathcal{C}_\tau^{(\gamma)}\label{c-exp}.
\end{align}
Computing the thermodynamic bound, Eq.~\eqref{tur-cauchy}, in this limit, we obtain:
\begin{equation}
\tau \geq \tau_0^{\rm bound} \left[1+ \left(\frac{\mathcal{L}^{(\gamma)}}{\mathcal{L}^{(0)}}-\frac{\mathcal{C_\tau^{(\gamma)}}}{2 C_\tau^{(0)}}  \right) \right] \equiv \tau_\gamma^{\rm bound},
\label{smallgamma}
\end{equation}
where $\tau_0^{\rm bound}=\mathcal{L}^{(0)}/\sqrt{2 \mathcal{C}_\tau^{(0)}}$ is the lower bound on the speed limit for the resetting-links free network. We emphasize that a similar relation can also be obtained even for a time-dependent unidirectional rate $\gamma(t)$ by setting its temporal average as a small parameter.

Eq.~\eqref{smallgamma} clearly shows that the thermodynamic bound on the speed limit in the presence of even one slow unidirectional transition can be tighter than $\tau_0^{\rm bound}$. Specifically, this occurs when:
\begin{equation}
\frac{\mathcal{L}^{(\gamma)}}{\mathcal{L}^{(0)}}~\geq~ \frac{\mathcal{C_\tau^{(\gamma)}}}{2 C_\tau^{(0)}},
\label{cond}
\end{equation}
which corresponds to the situation in which the relative statistical length increases more than the relative thermodynamic cost associated to it, due to the additional dissipation induced by the unidirectional link.

The advantage of performing a series expansion in powers of $\gamma$ is the possibility to isolate the extra contribution given solely by the energy dissipated through the unidirectional link. It is worth noting that the generalization to the case in which multiple unidirectional links are present is immediate, since the form of Eq.~\eqref{smallgamma}, and, consequently of Eq.~\eqref{cond}, does not change.

In Fig.~\eqref{fig:new-tur} we study a simple four-state network with one unidirectional link (indicated by a red one-head diagonal arrow), as shown in panel (a). In panel (b), we show $\tau_0^{\rm bound}$ (light green dashed curve), the bound including also the transition rate $\gamma$ (solid thin brown curve), and its approximation in the slow resetting regime $\tau_\gamma^{\rm bound}$ (dot-dashed blue curve). In this case we observe a tightening of the thermodynamic bound (see green thick diagonal line).

\section{Dynamical uncertainty relations for slow unidirectional transition rates}
\label{TUR}

This section is dedicated to the derivation of two inequalities and an uncertainty relation (TUR-like) for the (average) resetting entropy production, starting from the bounds derived in the previous sections. 

First of all, we notice that, since the system depends only on evolution time $t$, the statistical length and the thermodynamic cost function can be written as
\begin{eqnarray}
\label{Lp}
\mathcal{L} &=& \int_0^\tau~\mathrm{d}t~ \sqrt{\sum_{n=1}^{N}~\dfrac{(\partial_t p_n)^2}{p_n}}, \\
\mathcal{C} &=& \int_0^\tau~\mathrm{d}t~ \sum_{n=1}^{N}~\dfrac{(\partial_t p_n)^2}{p_n},
\label{Cp}
\end{eqnarray}
where we have used Eq.~\eqref{fmat}. This means that, even though $\mathcal{L}$ and $\mathcal{C}$ can be expressed in terms of thermodynamic quantities, in order to compute them, we just need information about the evolution of the probability distribution.

\begin{figure*}[t]
    \begin{center}
      \includegraphics[width=2 \columnwidth]{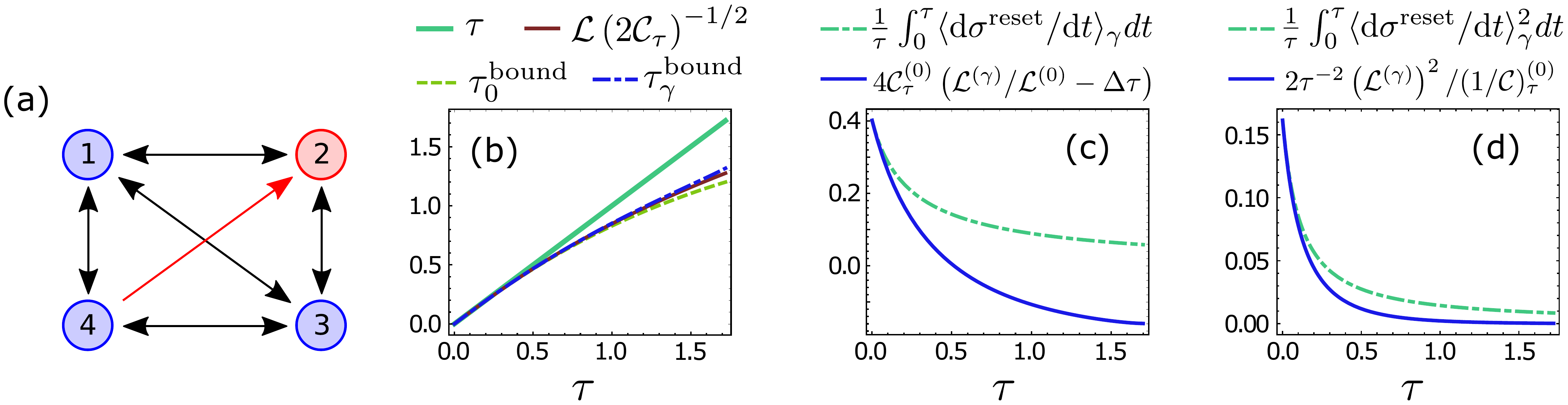}
      \caption{Thermodynamic bound on the speed limit for a system with one (slow) unidirectional transition. We consider the system as shown in panel a). Here, we initialize the system as $(p_1(0), p_2(0), p_3(0), p_4(0))=(0.7, 0.025, 0.15, 0.125)$. The unidirectional rate (in red) is $\gamma=0.1$, while the bidirectional one (in black) are: $W_{1 \to 2} = 1$, $W_{2 \to 1} = 0.5$, $W_{2 \to 3} = 0.48, W_{3 \to 2} = 0.4, W_{1 \to 3} = W_{3 \to 1} = 0.5$, $W_{3 \to 4} = 0.8, W_{4 \to 3} = 0.5$, $W_{1 \to 4} = 0.8$, and $W_{4 \to 1} = 0.4$. b) We show how the bound on the speed limit is modified by the presence of $\gamma$ (brown solid line), and the validity of the small-$\gamma$ approximation (blue dot-dashed line). In panels c) and d) we show, respectively, that the inequalities in Eqs.~\eqref{average} and \eqref{second} are satisfied. We remark that the maximum time $\tau$ for which we can study the system employing the small-$\gamma$ approximation becomes larger as $\gamma$ decreases.}
      \label{fig:new-tur}
    \end{center}
  \end{figure*}

Consider again the case in which a set of slow unidirectional links are added to a bidirectional backbone. By inverting Eq.~\eqref{smallgamma}:
\begin{gather}
\Delta \tau \geq\dfrac{\mathcal{L}^{(\gamma)}}{\mathcal{L}^{(0)}}-\dfrac{\mathcal{C}_\tau^{(\gamma)}}{2 \mathcal{C}_\tau^{(0)}},
\end{gather}
which results in
\begin{eqnarray}
\frac{1}{\tau} \int_0^\tau \mathrm{d}t \left\langle \frac{\mathrm{d}\sigma^{\rm reset}}{\mathrm{d}t} \right\rangle_\gamma \geq 4\mathcal{C}_\tau^{(0)} \left( \frac{\mathcal{L}^{(\gamma)}}{\mathcal{L}^{(0)}} - \Delta \tau \right).
\label{average}
\end{eqnarray}
where $\Delta \tau = (\tau - \tau_0^{\rm bound})/\tau_0^{\rm bound}$ is the relative deviation from the bound without resetting. Defining, for simplicity, $\mathbf{E}(\dot\sigma^{\rm reset}) = 2 \mathcal{C}^{(\gamma)}_\tau$, we have [see Eq.~\eqref{average}]:
\begin{equation}
\mathbf{E}(\dot\sigma^{\rm reset}) \geq 4\mathcal{C}_\tau^{(0)} \left( \frac{\mathcal{L}}{\mathcal{L}^{(0)}} - 1 - \Delta \tau \right).
\label{average-2}
\end{equation}
This is the first inequality involving the \textit{temporal} mean of $\langle {\rm d}\sigma^{\rm reset}/{\rm d}t \rangle_\gamma$, $\mathbf{E}(\dot\sigma^{\rm reset})$. In Fig.~\ref{fig:new-tur}(c), we numerically check the validity of Eq.~\eqref{average} in a pedagogical system.

A different inequality can be derived by applying the following Cauchy-Schwartz inequality:
\begin{equation}
\left( \int_0^\tau \frac{| \langle \dot\sigma^{\rm reset} \rangle_\gamma |}{2 \sqrt{\frac{{\rm d}s^2}{{\rm d}t^2}\big|_0}} \right)^2 \leq \frac{\tau}{2} (1/\mathcal{C})_\tau^{(0)} \int_0^\tau {\rm d}t \bigg| \left\langle \frac{{\rm d}\sigma^{\rm reset}}{{\rm d}t} \right\rangle_\gamma \bigg|^2,
\label{rel1}
\end{equation}
where the term inside the parenthesis on the left-hand side can be written as $\mathcal{L}^{(\gamma)}(|\langle \dot\sigma^{\rm reset} \rangle|) \equiv \mathcal{L}_{\rm abs}^{(\gamma)}$, and
\begin{equation}
(1/\mathcal{C})_\tau^{(0)} = \dfrac{1}{2\tau}\int_0^\tau {\rm d}t \left( \frac{{\rm d}s^2}{{\rm d}t^2}\bigg|_0 \right)^{-1}.
\end{equation}
Here, the quantity $\mathcal{L}^{(\gamma)}_{\rm abs}$ represents the perturbation to the statistical length due to $\gamma$, independently of the fact the the resetting link is producing entropy by erasing ($\dot{\sigma}^{\rm reset} < 0$) or writing ($\dot{\sigma}^{\rm reset} > 0$) information \cite{Busiello_2020}. Further, rearranging terms in Eq. \eqref{rel1}, we obtain:
\begin{gather}
\frac{1}{\tau} \int_0^\tau \mathrm{d}t \bigg| \left\langle \frac{\mathrm{d}\sigma^{\rm reset}}{\mathrm{d}t} \right\rangle_\gamma \bigg|^2 \geq \frac{2}{\tau^2} \left[ \frac{\left( \mathcal{L}^{(\gamma)}_{\rm abs} \right)^2}{(1/C)_\tau^{(0)}} \right].
\label{second}
\end{gather}
The above equation indicates that when there is at least one (slow) unidirectional transition rates, the second \textit{temporal} moment cannot be as small as possible, being bounded from below. Moreover, as expected, the faster is the process, the larger is its minimum value. In Fig.~\ref{fig:new-tur}(d), Eq.~\eqref{second} is numerically verified.

In order to have a compact notation, we rename the term on the left-hand side of Eq.~\eqref{second} as $\mathcal{V}\left(|\langle \dot\sigma^{\rm reset} \rangle|\right) \equiv \mathcal{V}_{\rm reset}$. Let us consider the following inequality:
\begin{gather}
\mathcal{L}^{(\gamma)}_{\rm abs} \mathcal{L}^{(0)} \geq \frac{\tau^2}{2} \Bigg(\frac{1}{\tau} \int_0^\tau dt \sqrt{\bigg| \left\langle \frac{{\rm d}\sigma^{\rm reset}}{{\rm d}t} \right\rangle_\gamma \bigg|} ~\Bigg)^2 = \nonumber \\
= \frac{\tau^2}{2} \mathbf{E}\left(\sqrt{|\dot\sigma^{\rm reset}|}\right)^2 \equiv \frac{\tau^2}{2} \mathcal{E}_{\rm reset},\label{eq-145}
\end{gather}
where we have used Cauchy-Schwarz inequality. 

Finally, using Eqs.~\eqref{rel1} and \eqref{eq-145}, we write an uncertainty relation for the resetting entropy production:
\begin{equation}
\frac{\mathcal{V}_{\rm reset}}{\mathcal{E}_{\rm reset}^2} \geq \left( \frac{\tau}{\mathcal{T}_{0}^{\rm bound}} \right)^2,
\label{tur-like}
\end{equation}
where we have introduced a different characteristic time:
\begin{equation}
\mathcal{T}_0^{\rm bound} = \mathcal{L}^{(0)} \sqrt{2 \left( 1/C \right)_\tau^{(0)}} \geq \tau_0^{\rm bound}.
\end{equation}
We remark that $\mathcal{T}_0^{\rm bound}$ is spiritually analogous to the bound on the speed limit due to bidirectional transitions only, $\tau_0^{\rm bound} = \mathcal{L}^{(0)} / \sqrt{2 \mathcal{C}^{(0)}_\tau}$. While in the former we consider the integral of the inverse of the squared intrinsic speed, $({\rm d}s^2/{\rm d}t^2)^{-1}$, the latter depends on the integral of the squared intrinsic speed, ${\rm d}s^2/{\rm d}t^2$.

Eq.~\eqref{tur-like} poses a limit on how much the resetting entropy production has to fluctuate, if we aim at erasing ($\langle \dot\sigma^{\rm reset} \rangle < 0$) or writing ($\langle \dot\sigma^{\rm reset} \rangle > 0$) information at a given rate, exploiting unidirectional transitions \cite{Busiello_2020}. We highlight that the quantities appearing in this TUR-like inequality are related to the ones usually involved in analogous relationships (variance and mean), even though they are not the same. Notably, in the limit of slow transitions, the right-hand side of Eq.~\eqref{tur-like} depends solely on zeroth order quantities (without resetting).

An additional remark is that the time $\tau$ for which a system can be studied under the small-$\gamma$ approximation gets larger when the value of $\gamma$ decreases. In the system under investigation in Fig.~\eqref{fig:new-tur} the value of the resetting link is not particularly small, even if all inequalities still remain valid, as evidenced by simulations.

\begin{figure}[t]
    \begin{center}
      \includegraphics[width=0.8\columnwidth]{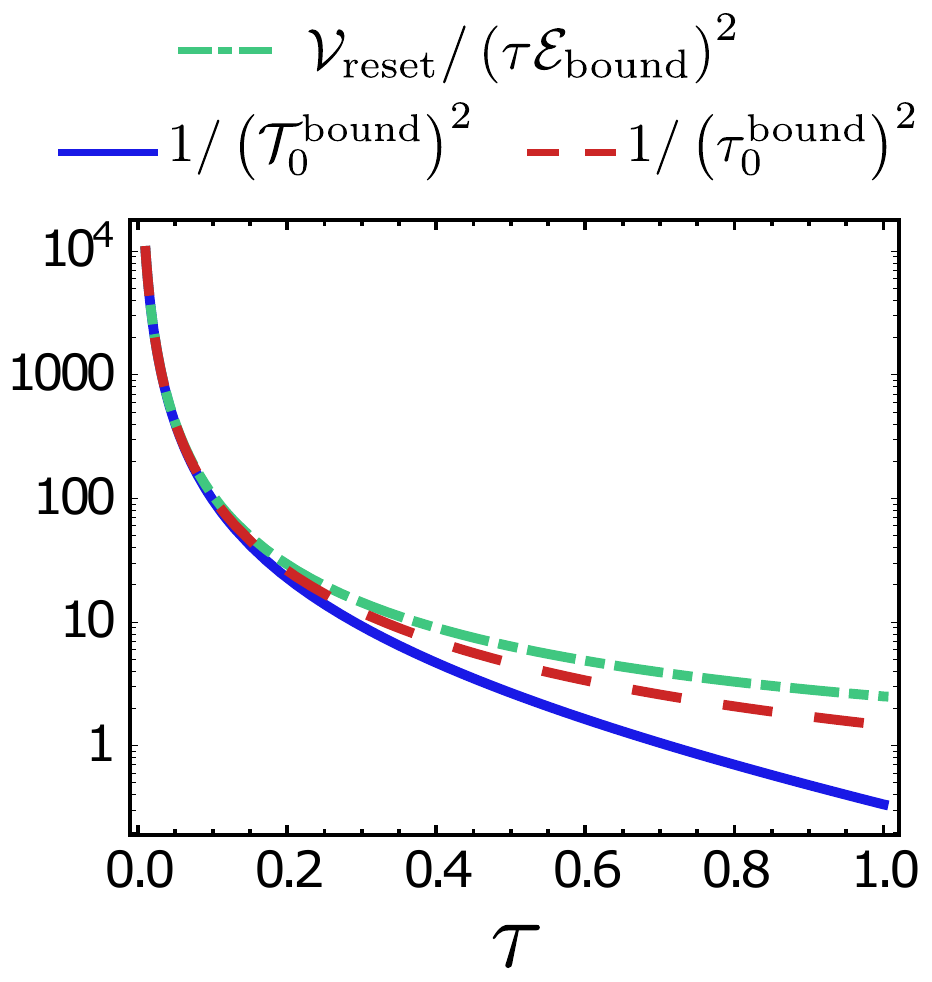}
      \caption{Thermodynamic uncertainty relation for the resetting entropy production [see Eqs.~\eqref{tur-like} and \eqref{tur-approx}]. The system under investigation is the same as in Fig.~\ref{fig:new-tur}(a). The green dot-dashed line is the left-hand side term of Eq.~\eqref{tur-like}, indicating the precision-to-cost ratio for resetting, divided by a factor $\tau^2$. The blue solid line represents $1/(\mathcal{T}_0^{\rm bound})^2$, which is the bound derived in the text. The red dashed  line is $1/(\tau_0^{\rm bound})^2$, indicating the bound in the small $\tau$ regime, Eq.~\eqref{tur-approx}. In this particular case, the latter term provides a tighter bound for the TUR-like inequality even for \textit{not-small} times.}
      \label{fig:TUR}
    \end{center}
  \end{figure}

A further refinement to the TUR-like inequality in Eq.~\eqref{tur-like} can be derived by considering the case of small $\tau$. In this limit, in fact, we have that $\mathcal{T}_0^{\rm bound} \to \tau_0^{\rm bound}$, which, in turns, implies:
\begin{equation}
\frac{\mathcal{V}_{\rm reset}}{\mathcal{E}_{\rm reset}^2} \gtrsim \left( \frac{\tau}{\tau_{0}^{\rm bound}} \right)^2.
\label{tur-approx}
\end{equation}
In Fig.~\ref{fig:TUR} we numerically show the validity of Eqs.~\eqref{tur-like} and \eqref{tur-approx} for the four-state system depicted in Fig.~\ref{fig:new-tur}(a). Numerical simulations evidences that this approximated version of the inequality provides a tighter bound to the precision-to-cost ratio even out of its range of validity, i.e., for \textit{not-small} times. We leave for future investigations a more in-depth study of this empirical observation.

\section{Thermodynamic bound on the speed limit}
\label{speed}
In this section, we obtain a \textit{new} bound on the speed at which the system goes from an initial to a final distributions in terms of thermodynamic cost function and statistical length. To this aim, we use the Milne’s inequality \cite{milne-1,milne-2,milne-3}:
\begin{align}
&\bigg[\int_0^\tau\mathrm{d}t~f(t)~g(t)\bigg]^2 \leq \int_0^\tau\mathrm{d}t_1~[f^2(t_1)+g^2(t_1)] \nonumber\\
&\times \int_0^\tau\mathrm{d}t_2~\dfrac{f^2(t_2)~g^2(t_2)}{f^2(t_2)+g^2(t_2)}\leq \int_0^\tau\mathrm{d}t_1~f^2(t_1)\int_0^\tau\mathrm{d}t_2~g^2(t_2),
\label{milne-ine}
\end{align}
to construct a relation between $\mathcal{L}$ and $\mathcal{C}$. Notice that, in principle, Milne's inequality may constitute a refinement of the Cauchy-Schwartz inequality. Proceeding as before, and substituting $g(t)=\phi$ (an arbitrary constant $\in \mathbb{R}$) and $f(t)=\frac{\mathrm{d}s}{\mathrm{d}t}$ in the above expression \eqref{milne-ine}, we get [see Eqs. \eqref{stln} and \eqref{costfn}]
\begin{align}
\mathcal{L}^2\leq (\phi^2 \tau+2 \mathcal{C})\bigg[\tau&-\overbrace{\phi^2\int_0^\tau~\mathrm{d}t~\bigg\{\phi^2+\frac{\mathrm{d}s^2}{\mathrm{d}t^2}\bigg\}^{-1}}^{\mathcal{M}(\tau,\phi)}\bigg]&\nonumber\\\nonumber\\&~~~~~~~~~~~~~~~~\leq 2 \tau \mathcal{C},
\label{tur-0}
\end{align}
where the equality is achieved when the square of the intrinsic speed $\frac{\mathrm{d}s^2}{\mathrm{d}t^2}$ is independent of time.
Notice that both integrands of $\mathcal{C}$ and $\mathcal{M}(\tau,\phi)$ contain the same term, which is the square of the intrinsic speed, $\frac{\mathrm{d}s^2}{\mathrm{d}t^2}$. The latter is related to thermodynamic quantities as displayed in Eq.~\eqref{fexp-ds2}.
\begin{figure*}
    \centering
    \includegraphics[width=2 \columnwidth]{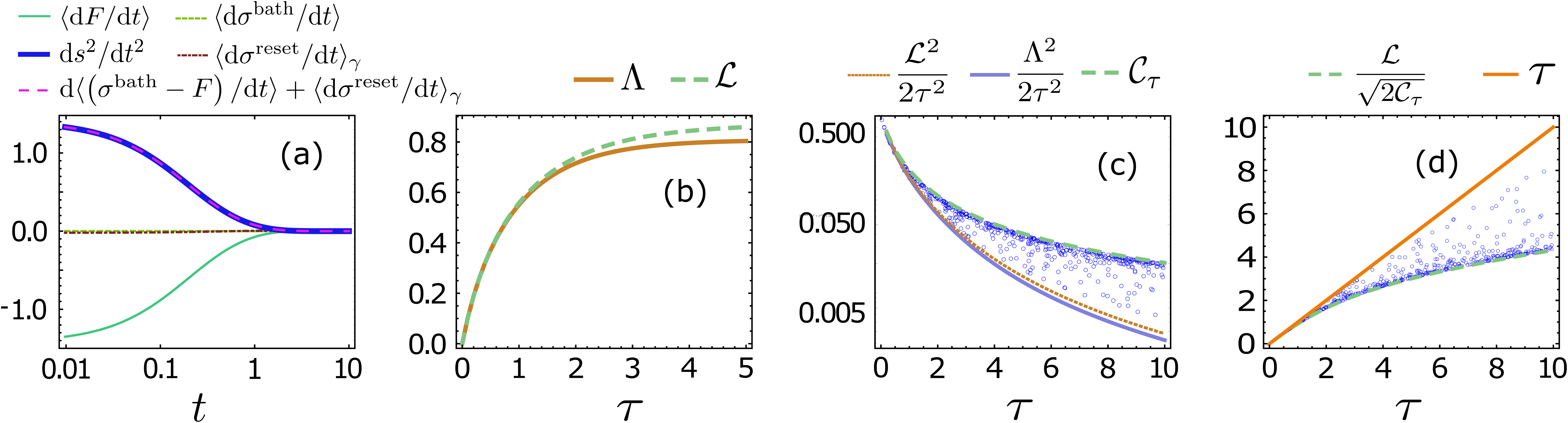}
    \caption{Thermodynamic bound on the speed limit for constant resetting rate. We consider the system as detailed in Fig.~\ref{fig:reset-1}. In panel (a), we show each component on the right-hand side of Eq.~\eqref{fexp-ds2}. Since bidirectional transitions are independent of time, $\big\langle \frac{\mathrm{d}\sigma^{\mathrm{bath}}}{\mathrm{d} t} \big\rangle$ is equal to zero for all time. Moreover, we show the verification of the Eq.~\eqref{fexp-ds2}, where the left-hand side is computed using Eq. \eqref{ds2-def}. In panel (b), we plot the inequality relating the statistical length and the shortest path between two states separated by time $\tau$: $\Lambda\leq \mathcal{L}$. Panel (c) demonstrates the inequality given in Eq.~\eqref{tur-3}, where the blue circles are obtained using the central term for random values of time $\tau\in[0,10]$ and $\phi\in[0,5]$ drawn from a uniform distribution. Clearly, the inequality obtained from first and second term could be tighter than the one derived using Cauchy-Schwartz inequality, presented in \cite{ito}, and involving first and third term of Eq.~\eqref{tur-3}. In panels (d), we show that the speed limit $\tau$ obeys a new inequality given in Eq.~\eqref{tur-f}. The numerical result hints at the possibility to find a tighter bound. Herein, blue circles are obtained from the second term given in Eq.~\eqref{tur-f} using random values of $\tau$ and $\phi$ [similar to panel (c)].}
    \label{fig:TUR-bound}
\end{figure*}

  \begin{figure*}
      \centering
      \includegraphics[width=2 \columnwidth]{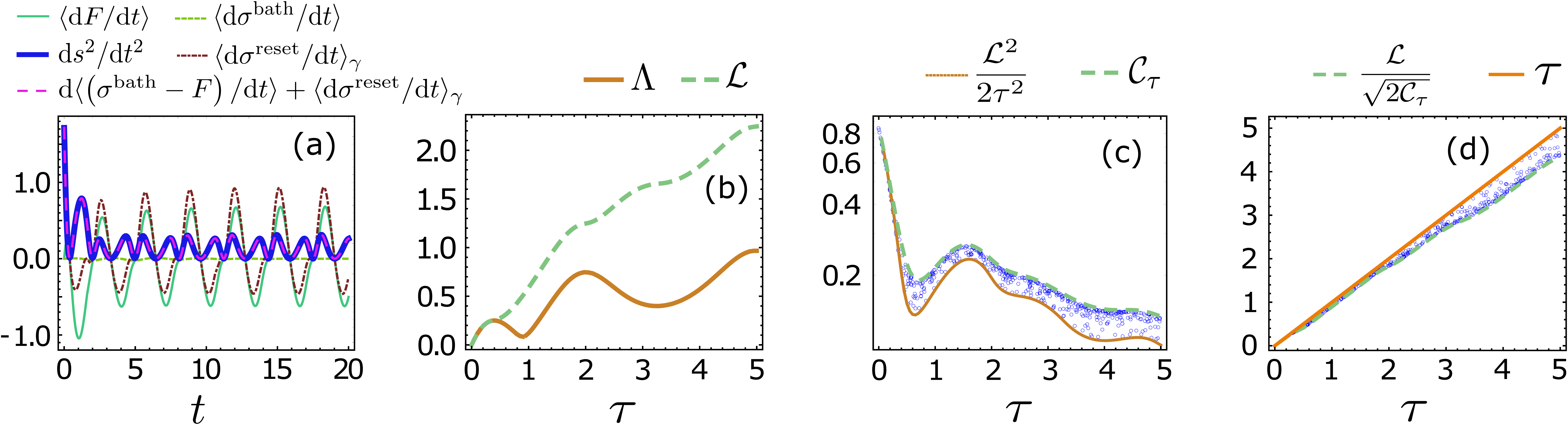}
      \caption{Thermodynamic bound on the speed limit for periodic resetting. We consider the system as shown in Fig.~\ref{fig:flux}(upper-panel). Here, we initialize the system as $(p_1(0), p_2(0), p_3(0), p_4(0))=(0.7, 0.025, 0.15, 0.125)$. The unidirectional and  bidirectional rates are $\gamma(t)=\gamma_n(t)=2 \cos^2(t), W_{2 \to 1}(t)=\tanh(0.5t), W_{1 \to 2}(t)=W_{2 \to 1}(0.3 t), W_{3 \to 2}(t)=\tanh(0.1t), W_{2 \to 3}(t)=W_{3 \to 2}(0.5t), W_{3 \to 1}(t)=W_{1 \to 3}(t)=\tanh(0.5t), W_{4 \to 3}=\tanh(0.5t), W_{3 \to 4}(t)=W_{4 \to 3}(0.2t), W_{1 \to 4}(t)=\tanh(0.5t), W_{4 \to 1}(t)=W_{1 \to 4}(0.5t)$. In analogy to Fig.~\ref{fig:TUR-bound}, in panel (a), we show each component of Eq.~\eqref{fexp-ds2}; in panel (b), we plot the inequality $\mathcal{L}\geq\Lambda$; finally, in panels (c) and (d), we verify the inequalities in Eqs.~\eqref{tur-4} and \eqref{tur-f}.}
      \label{fig:TUR-bound-2}
  \end{figure*}

In what follows, for convenience,  we will not write the time dependence in $\mathcal{M}(\tau,\phi)$. 
Rearranging the inequality \eqref{tur-0}, we obtain 
\begin{align}
\mathcal{C}_\tau\geq \dfrac{1}{2}[1-\mathcal{M}_\tau(\phi)](\phi^2+2\mathcal{C}_\tau)\geq \dfrac{\mathcal{L}^2}{2 \tau^2},
\label{tur-3}
\end{align}
Notice that $\mathcal{M}_\tau(\phi)$ is a non-null quantity.

Consider now the expressions of the statistical length $\mathcal{L}$ and the thermodynamic cost $\mathcal{C}$ in terms of the probability distribution, respectively Eqs.~\eqref{Lp} and \eqref{Cp}. These two functions can be simultaneously optimised with the constraint of the normalization of the probability distribution, by using the method of Lagrange multipliers. Following the steps presented in \cite{PRX}, the shortest path connecting two distributions in a finite time $\tau$, at the minimum cost, is
\begin{align}
\Lambda=2 \arccos\bigg(\sum_{n=1}^{N}\sqrt{p_n(0) p_n(\tau)}\bigg).
\label{short-len}
\end{align}
Therefore, the statistical length of any other path is always greater or equal to the shortest path at all times, i.e., $\mathcal{L}\geq \Lambda$. We remark that the optimal distribution $p^*_n(t)$, derived in \cite{PRX}, simultaneously minimizes $\mathcal{L}$ and $\mathcal{C}$, and let Eq.~\eqref{tur-0} holds as an equality. Indeed, when evaluated for $p_n^*(t)$, $\frac{\mathrm{d}s^2}{\mathrm{d}t^2}=\Lambda^2$, being constant in time, that is the condition for the equality to hold in Eq.~\eqref{tur-0}. 

Using the definition of $\Lambda$, the relation \eqref{tur-3} becomes 
\begin{align}
\mathcal{C}_\tau\geq \dfrac{1}{2}[1-\mathcal{M}_\tau(\phi)](\phi^2+2\mathcal{C}_\tau)\geq \dfrac{\mathcal{L}^2}{2 \tau^2}\geq \dfrac{\Lambda^2}{2 \tau^2},
\label{tur-4}
\end{align}
where the fourth term on the right-hand side is obtained using the shortest length $\Lambda$. Clearly, from first and last term, one can conclude that the (scaled) cost function is bounded from below by the net kinetic energy \cite{PRX} associated to the motion along the shortest path in the probability space in a time $\tau$, independently on how the path is traversed. Furthermore, using second, third and fourth terms of the inequality \eqref{tur-4}, we find a new uncertainty relation for $\mathcal{M}_\tau(\phi)$ and $\mathcal{C}_\tau$:  
\begin{align}
[1-\mathcal{M}_\tau(\phi)](\phi^2+2\mathcal{C}_\tau)\geq \dfrac{\mathcal{L}^2}{\tau^2}\geq \dfrac{\Lambda^2}{\tau^2}.
\label{tur-5}
\end{align}
Unfortunately, the physical interpretation of $\mathcal{M}_\tau(\phi)$ is not clear to us. However, as we showed earlier, this quantity contains the same term as in $\mathcal{C}_\tau$ [see Eq.~\eqref{tur-0}].  Nonetheless, from the above relation \eqref{tur-5}, we can say that for any physical system connecting two states in a finite time $\tau$, the product of two functions on the left-hand side will be always greater than or equal to the square of the speed of the system over the shortest path. We stress that the bound shown in Eq.~\eqref{tur-4} does not require that a system has both unidirectional and bidirectional links, but is also applicable to those with only bidirectional transitions.

Using Eq.~\eqref{tur-5}, we obtain the thermodynamic bound on the speed limit: 
\begin{align}
\tau&\geq \dfrac{\mathcal{L}}{\sqrt{[1-\mathcal{M}_\tau(\phi)](\phi^2+2\mathcal{C}_\tau)}}\nonumber\\&~~~~~~~~~~~~~\geq \dfrac{\Lambda}{\sqrt{[1-\mathcal{M}_\tau(\phi)](\phi^2+2\mathcal{C}_\tau)}}.
\label{tur-f}
\end{align}
Clearly, in the limit $\phi\to 0$, we reduce to the bound $\tau\geq \mathcal{L}/\sqrt{2\mathcal{C}_\tau}$, derived in the Ref.~\cite{ito}.

In Fig.~\ref{fig:TUR-bound}, we consider the system as discussed in Fig.~\ref{fig:reset-1}. We show the evolution of each component on the right-hand side of Eq.~\eqref{fexp-ds2} with respect to time in Fig.~\ref{fig:TUR-bound}(a). Therein, we choose all bidirectional transitions to be independent of time. Thus, the quantity $\big\langle \frac{\mathrm{d}\sigma^{\mathrm{bath}}}{\mathrm{d} t} \big\rangle$ remains zero at all times. We also verify Eq.~\eqref{fexp-ds2}, where the left-hand side is computed using Eq. \eqref{ds2-def}. In Fig.~\ref{fig:TUR-bound}(b), we confirm the inequality $\Lambda\leq \mathcal{L}$. Then, we show in Fig.~\ref{fig:TUR-bound}(c), the numerical evidence of the bound in Eq.~\eqref{tur-3}, indicating that the inequality obtained from first and second term could be tighter than the one derived using Cauchy-Schwartz inequality \cite{Cover}, and involving first and third term of Eq.~\eqref{tur-3}. The blue circles are obtained from the central term of the inequality~\eqref{tur-3}, using random values of time $\tau\in[0,10]$ and $\phi\in[0,5]$, drawn from a uniform distribution. Finally, in Fig.~\ref{fig:TUR-bound}(d), we plot the thermodynamic bound on the speed limit $\tau$ given in Eq.~\eqref{tur-f}, where the blue circles are obtained using the second term of the inequality \eqref{tur-f} in which $\tau\in[0,10]$ and $\phi\in[0,5]$ are drawn randomly from a uniform distribution. Clearly, the comparison between the bound on the speed limit previously obtained \cite{ito} and the one presented in Eq.~\eqref{tur-f} suggests that the latter could be tighter, depending on the value of $\phi$. Furthermore, we also perform the same analysis for the case when the resetting rate is a periodic function of time: $\gamma=\gamma_n=2 \cos^2(t)$. It is illustrated in Fig.~\ref{fig:TUR-bound-2}.
 
\section{Tighter bound}
\label{opt}
In this section, we obtain the tighter bound on the speed limit. In Figs.~\ref{fig:TUR-bound}(c)-(d) and \ref{fig:TUR-bound-2}(c)-(d), the blue circles are obtained from central terms of Eq.~\eqref{tur-4} and \eqref{tur-f}, where those terms are functions of $\phi$. 

\begin{figure}[!h]
    \begin{center}
      \includegraphics[width=4.3cm]{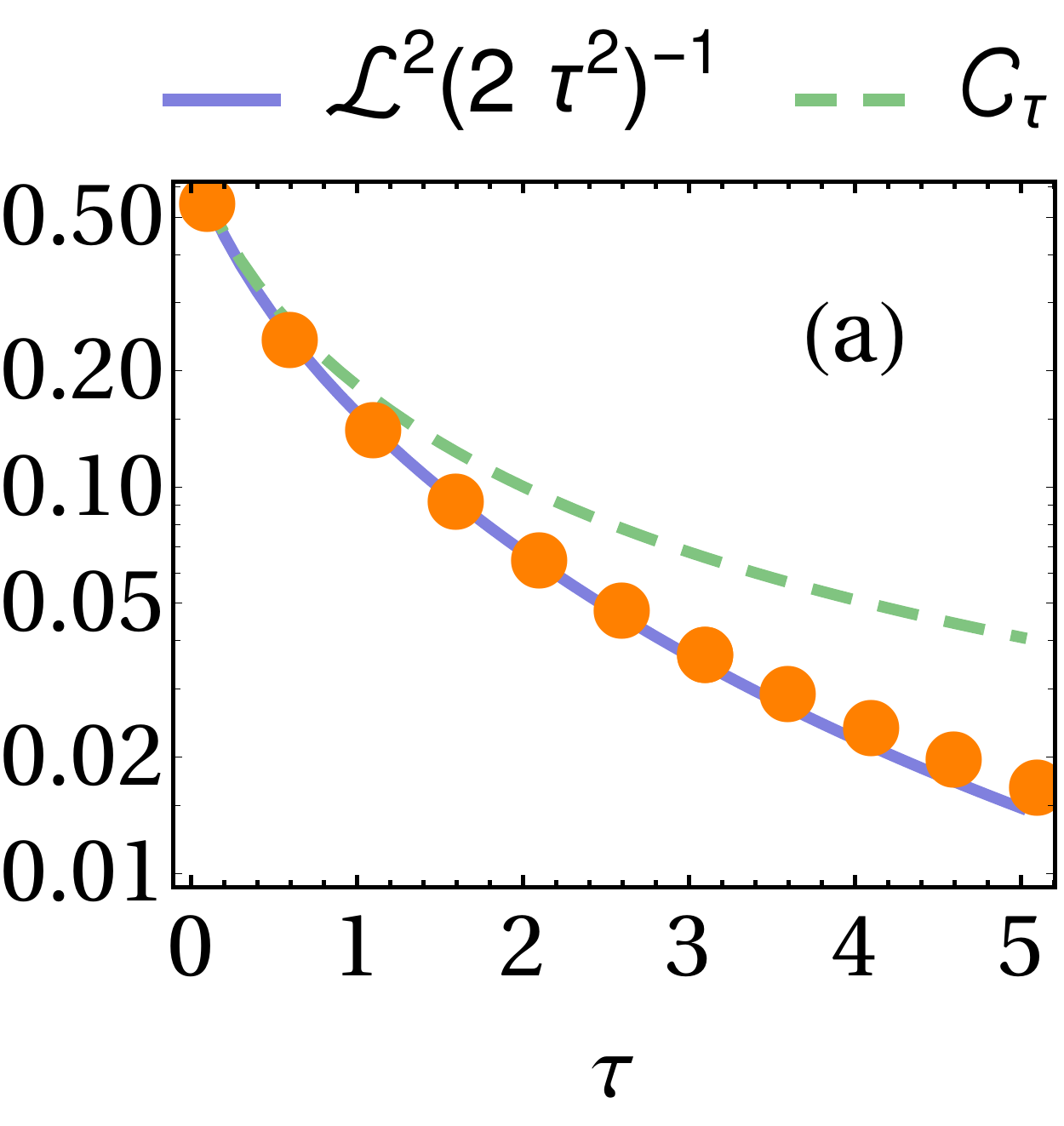}~~
      \includegraphics[width=4cm]{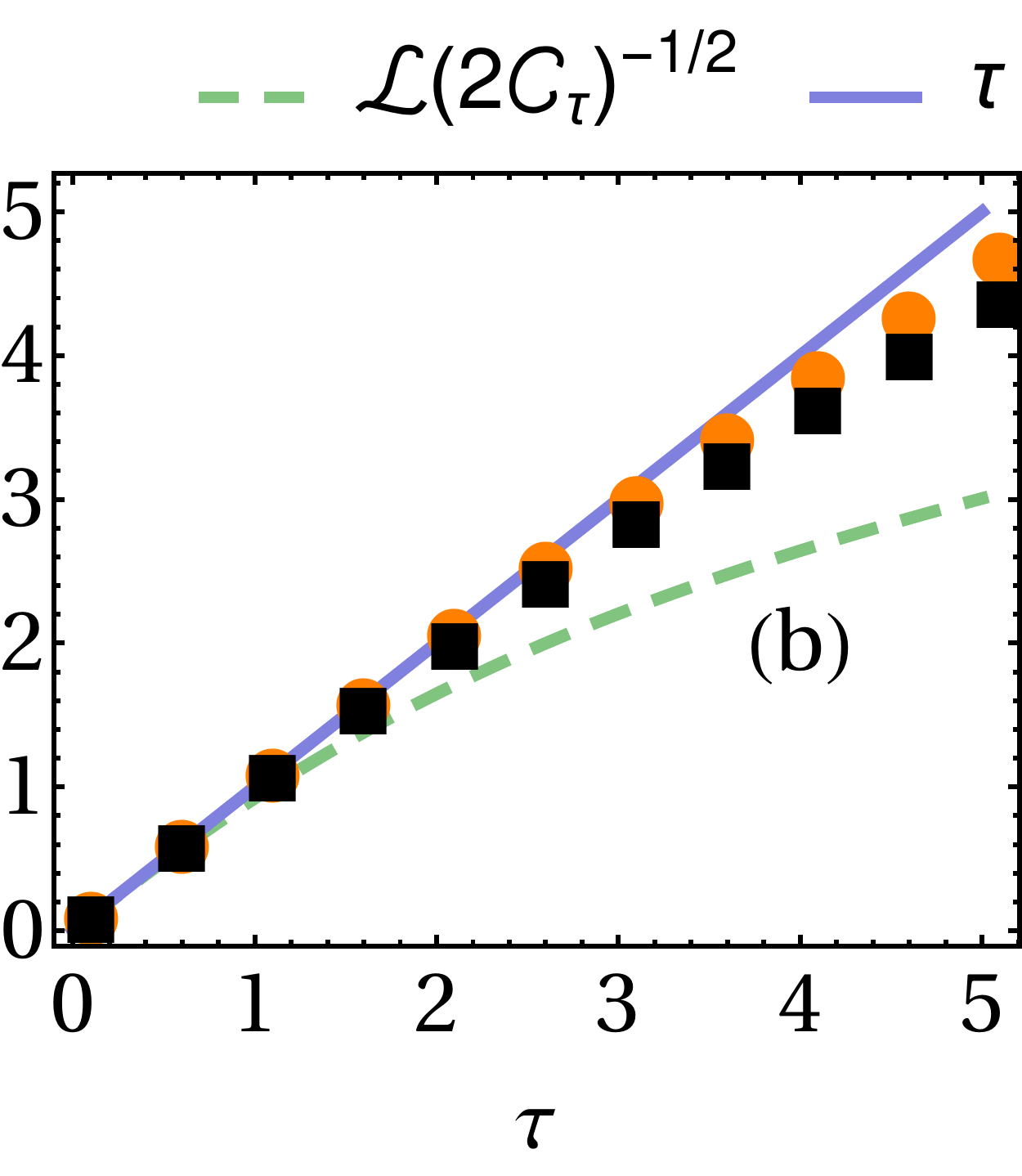}
            \caption{Tighter thermodynamic bound on speed limit. Here, we consider the system discussed in Fig.~\ref{fig:reset-1}. (a): We show the inequality \eqref{tur-6}, where orange circles are given by central term. (b): A tighter bound on the speed limit $\tau$, where orange circles and black squares, respectively, are obtained from the central and and the last term of Eq.~\eqref{tur-f2}.  }
      \label{fig:TUR-bound-3}
    \end{center}
  \end{figure}

Clearly, one can minimize the central term in Eq.~\eqref{tur-4}, obtaining the following tighter inequality [see Fig.~\ref{fig:TUR-bound-3}(a)]:
\begin{align}
\mathcal{C}_\tau\geq\min_{\phi}\bigg\{\dfrac{1}{2}[1-\mathcal{M}_\tau(\phi)](\phi^2+2\mathcal{C}_\tau)\bigg\}\geq \dfrac{\mathcal{L}^2}{2 \tau^2},
\label{tur-6}
\end{align}
Analogously, a tighter bound on the speed limit can also be derived by maximizing the terms on the right-hand sides of Eq.~\eqref{tur-f}:
\begin{align}
\tau&\geq \max_{\phi}\bigg\{\dfrac{\mathcal{L}}{\sqrt{[1-\mathcal{M}_\tau(\phi)](\phi^2+2\mathcal{C}_\tau)}}\bigg\}\nonumber\\
&~~~~~~~~~~~~~\geq \max_{\phi}\bigg\{\dfrac{\Lambda}{\sqrt{[1-\mathcal{M}_\tau(\phi)](\phi^2+2\mathcal{C}_\tau)}}\bigg\}.
\label{tur-f2}
\end{align}

In Fig.~\ref{fig:TUR-bound-3}(b), we show that the speed limit $\tau$ is tightly bounded by the functions in Eq.~\eqref{tur-f2}, and this bound is stronger than the previous one presented in the literature \cite{ito} (see dashed and solid line).

It is worth to remark that the relations in Eqs.~\eqref{average} and \eqref{second} can, in principle, be improved by using the Milne's inequality, and employing the optimization scheme, instead of the Cauchy-Schwartz inequality. However, they might result to be particularly cumbersome, so we decided to not report them in this manuscript.\\

\section{Summary}
\label{summ}

In this paper, we considered a general discrete state-space system with both unidirectional and bidirectional transitions. It evolves from a given initial distribution, following a Master Equation. We computed the bound on the evolution time to reach a final distribution in terms of the statistical length and the thermodynamic cost. These are two thermodynamic quantities that can be written as a function of three contributions: 1) the total entropy production, 2) the environmental entropy productions, and 3) the unidirectional (or resetting) entropy production. In particular, the latter contains information regarding unidirectional transitions, whose contribution has been highlighted, in particular in the limit of slow unidirectional transitions. Two inequalities and an uncertainty relation (TUR-like) involving the average resetting entropy production have been also derived. Using the Milne's inequality \cite{milne-1,milne-2,milne-3}, we built a \textit{new} thermodynamic bound on $\tau$, which is the time to connect two distributions, parameterized by a real number $\phi$. By means of an optimization procedure, we showed that our bound is tighter than the one previously reported \cite{ito}. Finally, we numerically validated our findings, using simple four-state models that stochastically reset with constant and periodic resetting rates.

Beyond refinement of the thermodynamic bound on the speed limit, and the generalization of the formalism presented in \cite{ito}, rooted on the theory of information geometry, to the case in which unidirectional transitions are present, here, we discuss a possible future application of this study. Consider a discrete-state system in which some information is accessible, e.g. the probability distributions at different times, or dissipation. It is possible, at least in principle, to employ and/or modify the uncertainty relations here derived to detect the presence of unidirectional transitions in the system. The possibility to pinpoint unidirectional links, whose unnoticed presence might irreversibly affect the system, would allow preventing detrimental changes. The investigation is under progress, and we hope that such results might be crucial for system suffering from catastrophic events \cite{sandro} (e.g. social and ecological communities).

\begin{acknowledgments}
D. G. is supported by ``Excellence Project 2018'' of the Cariparo foundation. D. G. and D. M. B. are grateful to Amos Maritan for insightful discussions and the careful reading of the manuscript.
\end{acknowledgments}

\appendix
\section{System with multiple unidirectional links}
\label{ul-sec}
In this section, we consider a generalization of the system introduced in the main text, whose dynamics is described by Eq.~\eqref{dyna-1}. In the same fashion, the system has bidirectional links such that $W_{n \to m}$ ($W_{m \to n}$) is the transition rate from a state $n$ to a state $m$ ($m$ to $n$). In addition to that, it also has unidirectional links such that $Y_{n\to m}>0$ is the transition rate from $n$ to $m$, but there is no conjugate transition rate: $Y_{m\to n}=0$. Following Ref.~\cite{Busiello_2020}, the master equation of this system can be written as
\begin{align}
\dfrac{{\rm d}p_n}{dt}&=\sum_{m\neq n}^{N} [W_{m\to n}p_m(t)-W_{n\to m}p_n(t)]\nonumber\\&+\sum_{m\neq n}^{N}[Y_{m\to n}p_m(t)-Y_{n\to m}p_n(t)].
\label{dyn-first}
\end{align}
where implicitly either $Y_{m \to n} = 0$ or $Y_{n \to m} = 0$, as discussed above. Note that, according to Eq.~\eqref{dyn-first}, $Y_{n \to n} = - \sum_{m \neq n} Y_{n \to m}$. Further, we can recover the master equation \eqref{dyna-1} for a system resetting to a particular state $m=n_0$ (where we set $n_0=1$) by setting the transition rate $Y_{m\to n}=\gamma_m \delta_{n,n_0}$ and $Y_{n\to m}=\gamma_n \delta_{m,n_0}$ 

Using the same strategy shown in Eqs.~\eqref{sys-ent}--\eqref{eq1}, we obtain the system entropy production: 
\begin{align}
\dot S^{{\rm sys}}=\sum_{n,m} W_{n\to m}p_n \log\dfrac{p_n}{p_m}+\sum_{n,m} Y_{n\to m}p_n \log\dfrac{p_n}{p_m}, \label{eq7}
\end{align}  
where the dot again represents the derivative with respect to time. Rewriting the above equation in the spirit of Eq.~\eqref{eq2}, we get
\begin{align}
\dot S^{{\rm sys}} &=& \overbrace{\dfrac{1}{2}\sum_{n,m} J_{n\to m}~F_{n\to m}}^{\dot S^{{\rm tot}}}- \overbrace{\dfrac{1}{2}\sum_{n,m} J_{n\to m}~\sigma_{n\to m}^{\rm bath}}^{\dot S^{\rm bath}} + \nonumber \\
&\;& - \underbrace{\sum_{n,m} J^{{\rm uni}}_{n\to m}~\sigma_{n\to m}^{\rm{uni}}}_{\dot{S}^{\rm uni}},
\label{parts}
\end{align}
where the quantities in the first line of the above equation are defined in Eqs.~\eqref{cj}--\eqref{sgj}, and the ones in the second line are
\begin{align}
J^{{\rm uni}}_{n\to m}&=Y_{n\to m}~p_n,\label{qu1}\\   
\sigma_{n\to m}^{\rm{uni}}&=\ln\dfrac{p_m}{p_n}\label{qu2},
\end{align}
which respectively, are the current and the entropy change due to the unidirectional transition (indicated by the superscript `uni') from $n$ to $m$. When $m=1$ is a state where the system is allowed to reset, as for the case in the main text, the quantities in Eqs.~\eqref{qu1} and \eqref{qu2} can be easily replaced by Eq.~\eqref{rj} and Eq.~\eqref{s-reset}, respectively.

\section{Fisher information metric: Unidirectional links}
\label{FIM}
In the following, we compute the Fisher information metric for the system discussed in Appendix~\ref{ul-sec}. To do so, we recall Eq.~\eqref{ds2-def}:
\begin{align}
    \dfrac{\mathrm{d}s^2}{\mathrm{d}t^2}=-\sum_{n=1}^{N} p_n \dfrac{\mathrm{d}}{\mathrm{d}t}\bigg(\dfrac{1}{p_n}\dfrac{\mathrm{d} p_n}{\mathrm{d}t}\bigg).
\end{align}
We substitute the dynamics \eqref{dyn-first} in the above equation and get:
\begin{align}
     \dfrac{\mathrm{d}s^2}{\mathrm{d}t^2}&=\sum_{m,n} p_n \dfrac{{\rm}d}{{\rm d}t}\bigg[-W_{n\to m}e^{-F_{n\to m}}\bigg]\nonumber\\
     &-\sum_{m,n}p_n \dfrac{{\rm}d}{{\rm d}t}\bigg[Y_{m\to n}\dfrac{p_m}{p_n}\bigg],
\end{align}
where the first term is given by Eq.~\eqref{fpart}. Let us now evaluate the second term:
\begin{align*}
    \sum_{m,n}p_n \dfrac{{\rm d}}{{\rm d}t}\bigg[Y_{m\to n}\dfrac{p_m}{p_n}\bigg]&=\sum_{m,n}\bigg[p_m \dfrac{{\rm d}Y_{m\to n}}{{\rm d}t}\nonumber\\
    &+p_n Y_{m\to n}\dfrac{{\rm d}}{{\rm d}t}\bigg(\dfrac{p_m}{p_n}\bigg)\bigg] = 
    \end{align*}
    \begin{align}
    &=\sum_{m,n}\bigg[p_m \dfrac{{\rm d}Y_{m\to n}}{{\rm d}t}\nonumber\\
    &-p_m Y_{m\to n}\dfrac{{\rm d}}{{\rm d}t}\bigg(\ln \dfrac{p_n}{p_m}\bigg)\bigg],
\end{align}
where the first term on the right hand side is zero since $\sum_n Y_{m\to n}=0$ (considering the diagonal part), and the second term using Eqs.~\eqref{qu1} and \eqref{qu2} can be written as
\begin{align}
    \sum_{m,n}p_n \dfrac{{\rm d}}{{\rm d}t}\bigg[Y_{m\to n}\dfrac{p_m}{p_n}\bigg]&=-\sum_{m,n} J^{{\rm uni}}_{m\to n}\dfrac{{\rm d} \sigma_{m\to n}^{\rm{uni}}}{{\rm d}t}\nonumber\\
    &=-\bigg\langle \dfrac{{\rm d} \sigma^{\rm{uni}}}{{\rm d}t} \bigg\rangle_{\gamma},\label{funi}
\end{align}
where the angular brackets represents the average over unidirectional trajectories.

Therefore, using Eq.~\eqref{fpart} and \eqref{funi}, we write the Fisher information metric as
\begin{align}
\dfrac{\mathrm{d}s^2}{\mathrm{d}t^2}&=\bigg\langle\dfrac{\mathrm{d}\sigma^{\mathrm{bath}}}{\mathrm{d}t}\bigg\rangle-\bigg\langle\dfrac{\mathrm{d}F}{\mathrm{d}t}\bigg\rangle+\bigg\langle\dfrac{\mathrm{d}\sigma^{\mathrm{uni}}}{\mathrm{d}t}\bigg\rangle_{\gamma},
\label{fexp-ds2-2}
\end{align}
which is an expression formally identical to the one derived for the case presented in the main text.

\end{document}